\documentclass[a4paper,12pt,man,natbib]{apa6}
\usepackage[utf8]{inputenc}
\usepackage{amsmath} 
\usepackage{amsfonts}
\usepackage{amssymb} 
\usepackage{graphicx}
\usepackage{float} 
\usepackage{breakcites}
\usepackage{pdfpages}
\usepackage{xcolor}
\usepackage{multicol}
\usepackage{natbib}
\usepackage{lineno}
\usepackage{caption}
\usepackage{subcaption}
\usepackage[ruled,vlined]{algorithm2e}

\makeatletter
\def\BState{\State\hskip-\ALG@thistlm}
\newcommand\bzero{\mbox{\boldmath${0}$}}
\newcommand\bbe{\mbox{\boldmath${ \beta}$}}

\newcommand\bfeta{\mbox{\boldmath${\eta}$}}
\newcommand\bep{\mbox{\boldmath${\epsilon}$}}

\newcommand\bdel{\mbox{\boldmath${\delta}$}}
\newcommand\bmu{\mbox{\boldmath${\mu}$}}

\newcommand\bnu{\mbox{\boldmath${\nu}$}}

\newcommand\bxi{\mbox{\boldmath${\xi}$}}

\newcommand\bSig{\mbox{\boldmath${\Sigma}$}}

\newcommand\btheta{\mbox{\boldmath${\theta}$}}
\newcommand\bTheta{\mbox{\boldmath${\Theta}$}}
\newcommand\bOmega{\mbox{\boldmath${\Omega}$}}

\newcommand\bB{{\bf B}}
\newcommand\bb{{\bf b}}

\newcommand\bD{{\bf D}}

\newcommand\be{{\bf e}}

\newcommand\bH{{\bf H}}

\newcommand\bI{{\bf I}}

\newcommand\bK{{\bf K}}
\newcommand\bk{{\bf k}}

\newcommand\mcN{{\mathcal N}}

\newcommand\mcS{{\mathcal S}}
\newcommand\bs{{\bf s}}
\newcommand\bt{{\bf t}}

\newcommand\bX{{\bf X}}

\newcommand\bY{{\bf Y}}

\newcommand\bZ{{\bf Z}}

\newcommand\mbR{{\mathbb R}}

\newcommand\bOne{{\bf 1}}

\makeatletter
\def\BState{\State\hskip-\ALG@thistlm}
\makeatother
\graphicspath{{Figures/}}

\title{Flood hazard model calibration using multiresolution model output}
\shorttitle{Flood hazard model multiresolution calibration}

\author{
Samantha M. Roth$^1$,\\ Ben Seiyon Lee$^2$,\\ Sanjib Sharma$^3$,\\ Iman Hosseini-Shakib$^3$,\\ Klaus Keller$^4$,\\ and Murali Haran$^1$
}
\affiliation{
Department of Statistics, Pennsylvania State University$^1$,\\ Department of Statistics, George Mason University$^2$,\\ Earth and Environmental Systems Institute, Pennsylvania State University$^3$, \\ Thayer School of Engineering at Dartmouth College$^4$
}

\authornote{\noindent Author Contacts: \\
 \begin{small} \noindent Samantha Roth: svr5482@psu.edu, Ben Seiyon Lee: slee287@gmu.edu, Sanjib Sharma: svs6308@psu.edu, Iman Hosseini-Shakib: ikh5084@psu.edu, Klaus Keller: klaus.keller@dartmouth.edu, and Murali Haran: muh10@psu.edu
 \end{small}
    }

\abstract{Riverine floods pose a considerable risk to many communities. Improving flood hazard projections has the potential to inform the design and implementation of flood risk management strategies. Current flood hazard projections are uncertain, especially due to uncertain model parameters. Calibration methods use observations to quantify model parameter uncertainty. With limited computational resources, researchers typically calibrate models using either relatively few expensive model runs at high spatial resolutions or many cheaper runs at lower spatial resolutions. This leads to an open question: Is it possible to effectively combine information from the high and low resolution model runs? We propose a Bayesian emulation-calibration approach that assimilates model outputs and observations at multiple resolutions. As a case study for a riverine community in Pennsylvania, we demonstrate our approach using the LISFLOOD-FP flood hazard model. The multiresolution approach results in improved parameter inference over the single resolution approach in multiple scenarios. Results vary based on the parameter values and the number of available models runs. Our method is general and can be used to calibrate other high dimensional computer models to improve projections.}


\begin{document}
\maketitle







\section{1. Introduction}\label{Sec:Intro}

Riverine flooding occurs when a river or stream exceeds its channel and flows onto the surrounding low-lying land \citep{FEMA_riverine_flooding, EPA_flood_risk}. Under the current climate the total expected annual damage from riverine flooding is about \$7 billion in the United States \citep{wobus_etal_2021}. These impacts are expected to increase with climate change and rapid urbanization \citep{IPCCreport, bates2021}. Flood hazard is dynamic and deeply uncertain \citep{zarekarizi_2020}. It is critical to characterize the uncertainties surrounding flood hazard estimates to inform the design of risk management strategies. 

Improved flood projections and uncertainty characterizations can help inform decisions to prepare for future flooding. Better flood projections can, for example, help homeowners to know if they should consider elevating their house or purchasing flood insurance \citep{zarekarizi_2020}. Projections for natural events are achieved through computer models or simulators, which are mathematical models that approximate the considered events and related systems. Such computer models are widely used, for instance, to project wildfires \citep{firemodel}, floods \citep{lisflood}, and droughts \citep{droughtmodel}. 



Computer models use parameter settings and other fixed inputs to produce features of the physical system of interest. Parameter settings can be used to represent uncertain quantities and physical processes. Fixed inputs can include spatial and temporal resolution. For example, in a flood hazard model the river discharge might also be treated as fixed, while other physical quantities that characterize the river might be treated as uncertain parameters. The goal of calibration is to adjust the values of the uncertain parameters to produce model outputs or projections that are most compatible with observations \citep{kennedy2001bayesian}.

There are a variety of both frequentist and Bayesian calibration approaches available \citep[cf.][]{cite:32,cite:30,cite:31,cite:29,kennedy2001bayesian,chang_binary,hall_etal_2011}. Calibration requires model runs at different parameter settings, and for complex physical models with longer run times, obtaining model runs at many parameter settings can impose nontrivial and sometimes prohibitive computational costs. Emulation-based approaches can help to reduce these computational burdens. Emulators, also called surrogate models, are computationally efficient approximations of the expensive computer model. Examples of emulators are Gaussian processes (cf. \cite{kennedy2001bayesian,cite:24,hall_etal_2011}), polynomial chaos expansions (cf. \cite{pce_liu2020, pce_slot2020}), and convolutional neural networks \citep{fletcher_multires}. Statistical emulators such as the Gaussian process emulator we develop here provide uncertainties and probability distributions to go along with the approximations. This is valuable for developing a rigorous computer model calibration approach.




Limited computational resources often leads to a trade-off between either using many lower resolution model runs or fewer higher resolution model runs for calibration. 
Approaches exist to use multiple resolutions of model runs in emulation \citep{fletcher_multires,jck_ml,cite:13}, but to our knowledge none have been used in emulation-calibration. 
In this manuscript, we focus on Gaussian process-based emulators \citep{jck_ml, cite:13} which are flexible and easy to use, and we build upon the multiresolution Gaussian process emulator of \cite{jck_ml}. In an example using an offshore wind farm computer model, \cite{jck_ml} show that their multiresolution Gaussian process emulator outperforms a Gaussian process emulator that uses only high resolution model output. We develop a multiresolution emulation-calibration approach to calibrate a flood hazard model. In an example using the LISFLOOD-FP flood hazard model for Selinsgrove (PA), our multiresolution Gaussian process emulation-calibration provides varying results that depend on the true parameter settings and the number of available model runs.

The remainder of this paper is organized as follows. In Section 2, we describe the computer model. In Section 3.1, we provide an overview of computer model calibration and existing emulation-calibration methods. We propose our multiresolution calibration approach in Section 3.2. Section 4 provides a numerical study and application to the LISFLOOD-FP model. We provide a summary and discuss avenues for future research in Section 5. 

\section{2. Data and Flood Hazard Model Description}\label{Sec: FloodHazardModel}

In this section we provide background information for the deterministic LISFLOOD-FP flood hazard model \citep{lisflood}. We demonstrate our methodology in the context of a case study in Selinsgrove, a riverine community in the Susquehanna River Valley in Pennsylvania. The output from the model is flood depth (the height of the water above the ground in meters) across the spatial domain of interest. 

The LISFLOOD-FP hydraulic model has been widely used to project flood hazards at a wider range of spatial and temporal scales \citep[cf.][]{FSF,regionalFloodRisk,Rajib_etal_2020,LisfloodCongo}. We use LISFLOOD-FP with the subgrid formulation to simulate flood hazards \citep{lisflood}. LISFLOOD-FP is a 2D hydraulic model for subcritical flow that solves the local inertial form of the shallow water equations using a finite difference method on a staggered grid. The model requires input related to ground elevation data describing the floodplain topography, channel bathymetry information (river width, depth, and shape), and inflow to the modeling domain as the boundary condition information. To apply LISFLOOD-FP, we use the subgrid-scale hydrodynamic scheme of \cite{nealetal_2012} to solve the momentum and continuity equations for both channel and floodplain flow. The scheme operates on a rectangular grid mesh of the same resolution as the input digital elevation model (DEM), using a finite difference scheme to solve the governing equations. The cells’ water depths are updated using mass fluxes between cells while ensuring mass conservation. 

To configure LISFLOOD-FP, we use floodplain topography information from the Pennsylvania Spatial Data Access (PASDA) archive \citep{PASDA}. We run LISFLOOD-FP using a 10 meter (m) and 50 m digital elevation model (DEM) \citep{PASDA}. For a 10 m DEM, the size of a grid cell is $10 \times 10$ $\textnormal{m}^2$. We select these two resolutions to learn how much information a relatively low resolution version of the model can provide when calibrating the model at a relatively high resolution. There is a tradeoff between model resolution and the associated computational costs. At the coarser spatial resolution (50 m), a single model run has an average walltime of 16 seconds on on the Pennsylvania State University's ICDS Roar supercomputer. At the finer spatial resolution (10 m), a single model run  has an average walltime of 6.1 minutes. These times exclude when the model fails and must be re-run, which occurs more frequently at the higher resolution. Accordingly, we call the 10m resolution model runs `expensive' and the 50m resolution model runs `cheap.' We use the river discharge corresponding to the 2011 Tropical Storm Lee as the inflow boundary condition ($13110.7 \frac{m^3}{s}$) \citep{USGS_peakflow}. This was one of the largest flood events in Selinsgrove in recent decades \citep{gitro_etal_2014, sanjib_etal_2022}.

\begin{figure}
     \centering
     \begin{subfigure}[b]{0.48\textwidth}
         \centering
         \includegraphics[width=\textwidth]{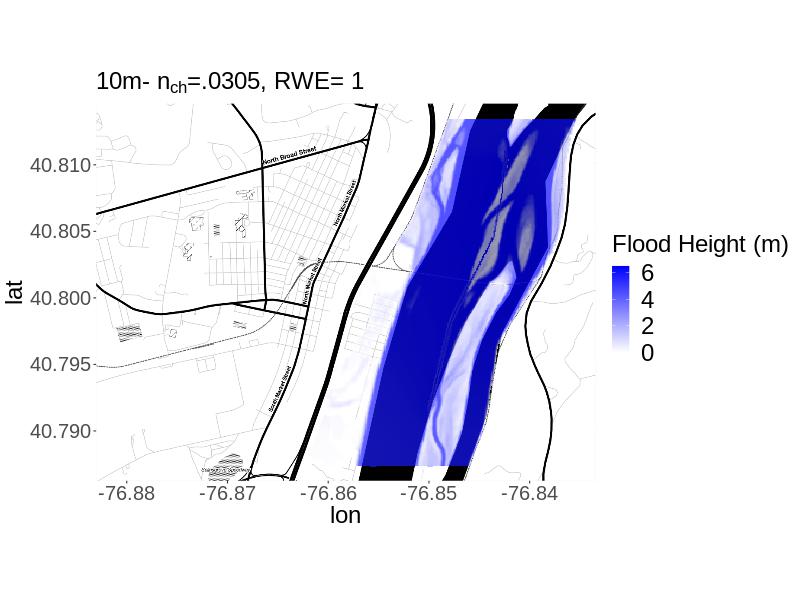}
         \label{fig:OG10mrun}
     \end{subfigure}
     \hfill
     \begin{subfigure}[b]{0.48\textwidth}
         \centering
         \includegraphics[width=\textwidth]{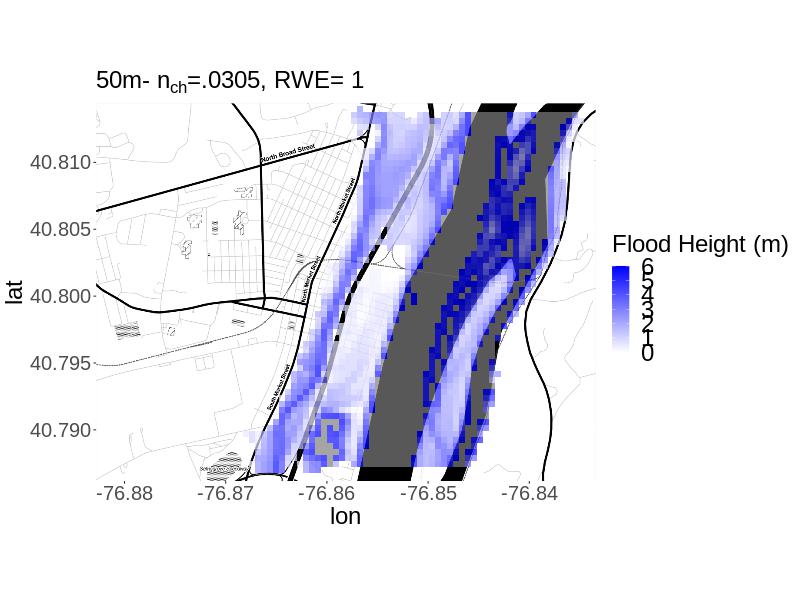}
         \label{fig:OG50mrun}
     \end{subfigure}
     \caption{LISFLOOD-FP model output at 10m resolution (left) versus 50m resolution (right) for the same parameter setting}
    \label{Fig: 10mVS50mSamePars1}
\end{figure}

We consider uncertainty underlying two parameters: (i) river width error ($RWE$), and (ii) Manning's roughness coefficient for the channel, e.g. channel roughness ($n_{ch}$). River width is estimated at different cross-sections of the river based on a map of the river and model runs. These estimates typically contain some errors. We apply the same multiplicative error term ($RWE$) to the river width estimates at all five cross-sections. Channel roughness estimates the resistance to flood flow in the channel \citep{arcement1989}. The ranges of plausible values determined by our own expert assessment for the parameters are $n_{ch} \in (0.02,0.1)$ and $RWE \in (0.95,1.05)$. For $n_{ch}$, expert assessment of the plausible range is informed by \cite{alipour_etal2022} and \cite{pappenberger_2008}. For $RWE$, expert assessment is informed by analyzing the ArcGIS basemap and applying domain knowledge of the realistic error range. To illustrate the difference between the LISFLOOD-FP model outputs at the 10m versus 50m resolution for the same parameter setting, we display the output for $n_{ch}=0.0305$ and $RWE=1$ (Figure \ref{Fig: 10mVS50mSamePars1}). For the same parameter setting, the lower resolution model run shows greater flood extent. 


\section{3. Model Calibration Framework}

In this section, we describe the general computer model calibration framework as well as the two-stage emulation-calibration approach. Then we introduce our multiresolution emulation-calibration method. 

\subsection{3.1 Computer Model Calibration}

In computer model calibration, key model parameters are inferred by comparing the associated computer model outputs to observational data \citep[cf.][]{chang_binary,kennedy2001bayesian,cite:27,cite:10,cite:29}. The computer model calibration framework is designed to account for important sources of uncertainty stemming from unknown input parameters, observational errors, and model-observation discrepancies. We define $\bY(\btheta,\bs)$ as the spatial computer model output, e.g. model run, at location $\bs$ = (latitude, longitude) $\in \mcS \subset \mbR^2$ and parameter setting, e.g. design point, $\btheta \in \bTheta$. $\mcS$ is the spatial domain of interest and $\bTheta$ is the parameter space. Here $\bTheta \subset \mbR^2$ since we wish to infer two parameters. $\bZ(\bs)$ is the observation at location $\bs$. We have access to computer model runs at $p$ design points, $\btheta_1,...,\btheta_p \in \bTheta$. Each model run  $\bY(\btheta_i)=$ $(Y(\btheta_i,s_1),...,  Y(\btheta_i,s_n))^T$ $\in \mbR^{n}$ at design point $\btheta_i$ is a spatial process observed at $n$ locations $\bs_1,..., \bs_n$. $\bZ= (Z(s_1),..., Z(s_n))^T \in \mbR^{n}$ is the vector of observations at these same locations. We treat the computer model as a `black box' and refrain from manipulating the internal mathematical models. 

In the Bayesian computer model calibration framework, the observation $\bZ$ is modeled as: 
\begin{align} \label{EQ:calibration}
    \bZ &= \bY(\btheta) + \bdel + \bep \nonumber, \\
    \bdel &\sim \mcN(\bzero,\bSig_{\bxi}), \textnormal{ and} \\
    \bep &\sim \mcN(\bzero,\sigma^2_\epsilon \bI) \nonumber
\end{align}
where $\bep$ is the independently and identically distributed observational error and $\bdel$ is a term used to characterize the systematic model-observation discrepancy. The discrepancy $\bdel$ is typically modeled as a zero-mean Gaussian process defined by spatial covariance matrix $\bSig_{\bxi}$ between the locations and the vector of covariance parameters $\bxi$. The discrepancy term is an essential component of model calibration 
\citep{discrepancy_importance}. To complete the Bayesian calibration framework, we specify prior distributions for $\btheta$, $\sigma^2_\epsilon$, and $\bxi$. 
Inference for $\btheta$, $\sigma^2_\epsilon$, and $\bxi$ is based on the posterior distribution $\pi(\btheta, \sigma^2_\epsilon, \bxi | \bZ)$, which can be approximated using samples drawn via a Markov Chain Monte Carlo (MCMC) algorithm. The aim is to identify $\btheta^{*}$, a most plausible parameter setting for the computer model and characterize and quantify the associated uncertainty. However, in reality the existence of a single best setting to link an imperfect computer model model with the physical system it represents may be unrealistic \citep{ReifiedBayesianModelling}.

The Bayesian calibration framework calls for a computer model run at each iteration of the MCMC algorithm. This approach can be computationally prohibitive for computer models with even moderately long run times on the order of minutes. For LISFLOOD-FP, 100,000 iterations from the Metropolis-Hastings algorithm, using an all-at-once update, would take approximately 423 days on the Pennsylvania State University's ICDS Roar supercomputer. We avoid this problem by constructing a Gaussian process emulator based on a training set of computer model runs. 

In this study, we focus on the two-stage emulation-calibration approach, which fits an emulator to the computationally-intensive computer model first (Equation \eqref{EQ:GPemulation}) and then calibrates the resulting emulator with respect to observed data (Equation \eqref{EQ:GPcalibration}). Single-stage methods \citep{higdon2008} combine the two stages (emulation and calibration) into a single inferential step. However, two-stage approaches have several advantages over single-stage methods such as computational efficiency, better diagnostics, and ``cutting feedback'' by preventing one part of the model from impacting the other  (see \cite{cite:12, cite:10,rougier2008comment,bayarri2007framework}).

We provide an overview of the two-stage emulation-calibration framework for a spatial computer model. 
In the emulation step we fit a Gaussian process emulator to training data. The training data is comprised of concatenated spatial computer model outputs $\mathbf{Y}= \big(\bY(\btheta_{1}),...,\bY(\btheta_{p})\big)^T$ evaluated at the p design points. 
The Gaussian process emulator $\bfeta(\btheta,\bY)$ is constructed by fitting the model:
\begin{align}\label{EQ:GPemulation}
    \bY \sim \mcN(h(\bX)\bbe,\Sigma_{\bxi_Y}(\bX) + \sigma^2 I)
\end{align}
\noindent where $\bX$ is a $np \times b$ matrix of covariates including the spatial locations and the computer model parameters. These covariates are used to define $\Sigma_{\bxi_Y}(\bX)$, and $\bxi_Y$ is the vector of covariance parameters. $h(\bX)$ is a function $h(\cdot)$ of the matrix of parameter settings and spatial locations $\bX$. $\bbe \in \mbR^b$ and $\sigma^2 \in \mbR$ are the regression coefficients and nugget parameters, respectively. We estimate $\bbe$, $\bxi_Y$, and $\sigma^2$ using maximum likelihood to fit a Gaussian random field to $\bY$. This field gives a probability model for the computer model run at any parameter setting $\btheta \in \bTheta$ and any location $s \in \mcS$. The Gaussian process model gives a predictive distribution for $\bY(\btheta^0)$ at unobserved $\btheta^0$ given the runs used for training $\bY$. The emulator $\bfeta(\btheta,\bY)$ is the resulting interpolated process.


In the calibration step, we model the observed data $\bZ$ with respect to the emulator $\eta(\bY,\btheta)$ as follows:
\begin{align} \label{EQ:GPcalibration}
    \bZ &= \eta(\bY,\btheta) + \bdel + \bep \nonumber, \\
    \bdel &\sim \mcN(\bzero,\bSig_{\bxi}), \textnormal{ and} \\
    \bep &\sim \mcN(\bzero,\sigma^2_\epsilon \bI) \nonumber
\end{align}
where the discrepancy term $\bdel$ represents the systematic differences between emulator projections (rather than the computer model projections) and observations \citep[cf.][]{chang2014}, and $\bdel \sim \mcN(\bzero,\bSig_{\bxi})$ is a spatial covariance matrix as before. $\bep$ again represents the observational errors \citep[cf.][]{chang2014}. Similar to the framework in \eqref{EQ:calibration}, we can infer $\btheta$, $\bdel$ and $\sigma^2_\epsilon$ by sampling from the posterior distribution $\pi(\btheta,\bdel,\sigma^2_\epsilon|\bZ)$ via MCMC.


\subsection{3.2 Multiresolution Gaussian Process Emulation-Calibration}\label{SubSec:Multiresolution}
Here we introduce a dimension-reduced approach to calibrate a computer model by combining information from high resolution (expensive) model runs and lower resolution (cheap) model runs. We first describe the dimension-reduced emulation-calibration framework (Section 3.2.1), and we then describe the multiresolution emulator (Section 3.2.2) replacing a traditional Gaussian process emulator in this framework. 

\subsubsection{3.2.1 Dimension-reduced Gaussian process emulation-calibration}\label{SubSec:DimReduce}

While our focus is calibrating an expensive computer model using model runs at multiple resolutions, we also address the computational challenges associated with calibrating computer models with high-dimensional spatial outputs. At the 50 m and 10 m resolutions, each model run contains observations at 14,214 locations and 126,791 locations, respectively. Although Gaussian process emulation-calibration can be useful for dealing with computationally prohibitive model run times \citep{hall_etal_2011}, high-dimensional model outputs pose computational challenges of their own \citep{katzfuss2020vecchia,chang2014}, particularly in training the emulator and generating emulator output. For example, fitting an emulator calls for a Cholesky decomposition of a dense positive definite matrix with $\mathcal{O}(\frac{1}{3}n^{3})$ floating-point operations (FLOPs); hence, emulation can be computationally expensive in high-dimensional settings. Many researchers have developed methods to address these challenges which are prevalent in environmental sciences \citep{cite:10,cite:26}. Methods to deal with emulation-calibration for high dimensional model runs include basis expansion approaches \citep{cite:26,higdon2008}, dimension reduction approaches \citep{chang_binary,chang2014,cite:27}, and composite likelihood approaches \citep{cite:25}. In this study, we extend the dimension-reduction approach of \cite{chang2014}.

Consider $\bY$, a matrix of concatenated expensive and cheap model runs such that $$\bY = \big((\bY^E)^T,(\bY^C)^T\big)^T=(\bY^E(\btheta^E_1),...,\bY^E (\btheta^E_{p^E}), \bY^C(\btheta^C_1),..., \bY^C(\btheta^C_{p^C})\big)^T,$$ with $C$ indicating cheap and $E$ indicating expensive. Here $p^E$ and $p^C$ are the numbers of expensive and cheap design points. We require cheap model runs at all of the same design points as the expensive model runs, i.e. $\{\btheta_1^E,..., \btheta^E_{p^E}\} \subset \{\btheta_1^C,..., \btheta^C_{p^C}\}$. Each design point contains a setting for channel roughness and river width error. So that $\bY^E$ and $\bY^C$ are observed at the same locations, we interpolate the cheap model runs onto the same locations as the expensive model runs using bilinear interpolation.

Following \cite{chang2014}, we first center the columns of $\bY$ then apply singular value decomposition to obtain the scaled eigenvectors, $\bk_1= \sqrt{\lambda_1}\be_1, ... , \bk_p= \sqrt{\lambda_1}\be_p$, where $\lambda_1 > ... > \lambda_p$ are the ordered eigenvalues and $\be_1,...\be_p$ are the corresponding eigenvectors of the sample covariance matrix of $\bY$. We select the leading $J_y$ principal components that explain 95\% of the variation, then we form the basis matrix $\bK_y = (\bk_1 ,..., \bk_{J_y} )$. Next, we compute the $p \times J_y$ dimension-reduced matrix of model runs as:
\begin{align}
    \bY^R = \bY \bK_y (\bK^T_y\bK_y)^{-1}.
\end{align}
where $\{\bY^R\}_{ij}$ corresponds to the $i$th design point and the $j$th principal component. 

Next, we fit a modified version of the multiresolution emulator of \cite{jck_ml} to each column of $\bY^R$. The emulator for all principal components $\bfeta(\bY^R,\btheta)$ is the collection of predictive processes for the $J_y$ principal components at $\btheta$. These predictive processes are defined by the mean (Equation \eqref{EQ:emulatorMean}), covariance function (Equation \eqref{EQ:emulatorCov}), and emulator parameter estimates detailed in Section 3.2.2. We project the value of an expensive computer model run at new design point $\btheta_0$ in the original space by computing $Y_0= \bK_y \bfeta(\bY^R,\btheta_0)$.

At the observation-level scale, the calibration (statistical) model can be represented as $\bZ = \bK_y \bfeta(\btheta,\bY^R) + \bK_d \bnu + \bep$, where the discrepancy $\bdel$ is replaced by its kernel convolution representation $\bK_d \bnu$ \citep{higdon1998} and $\bep$ is the usual observational error term. Here $\bnu \sim N(0,\kappa_d I_{J_d})$ with $J_d < n$, where $\kappa_d$ describes the magnitude of the discrepancy. More information on selection of the kernel basis $\bK_d$ can be found in \cite{chang2014}. 

In this study, we chose to leverage the low-dimensional space of the principal components as well as its computational benefits. Here, we model the dimension-reduced observations $\bZ^R = (\bK^T\bK)^{-1} \bK^T (\bZ- \bmu_\bY)$, where $\bmu_\bY$ is the vector of column means of $\bY$ and $\bK = ( \bK_y , \bK_d )$. Then the dimension-reduced probability model becomes
\begin{align}
    \bZ^{R} \sim N \Bigg( \begin{bmatrix} \bmu_{\bfeta} \\ \bzero \end{bmatrix}, \begin{bmatrix} \bSig_{\bfeta} & \bzero \\ \bzero & \kappa_d I_{J_{d}} \end{bmatrix} + \sigma^2 (\bK^T\bK)^{-1}\Bigg).
\end{align}
where $\bmu_{\bfeta}$ and $\bSig_{\bfeta}$ are the mean and covariance of $\bfeta(\btheta,\bY^R)$. In our case, $\bZ$ is an expensive computer model run contaminated by independent and identically distributed Gaussian noise. Then there is no need to model a discrepancy term, and the probability model for $\bZ^R = (\bK_y^T\bK_y)^{-1} \bK_y^T (Z- \bmu_\bY)$ becomes 

\begin{align}
    \bZ^R \sim N ( \bmu_{\bfeta} , \bSig_{\bfeta} + \sigma^2 (\bK_y^T\bK_y)^{-1}).
\end{align}


\subsubsection{3.2.2 Multiresolution Gaussian Process Emulation} \label{SubSubSec:MultiresEmulation}


We extend the multiresolution emulation approach from \cite{jck_ml} which was developed for a temporal stochastic computer model to the context of a spatial deterministic computer model. Our adaptation of their approach models the expensive computer model runs as homoscedastic rather than heteroscedastic. \cite{jck_ml}'s approach links the cheap and expensive computer model runs through a similar autoregressive structure to that of \cite{cite:13}.  \citet{cite:13} employs a multi-stage emulator training procedure that first estimates the emulator parameters for the cheap model then plugs in those estimates to the equation to estimate the emulator parameters for the expensive model. On the other hand, \citet{jck_ml} provides a way to infer all emulation parameters simultaneously by modeling the concatenated cheap and expensive model runs in one multivariate normal distribution as shown in Equation \eqref{EQ: JCK_MVN}. When training the cheap and expensive model emulators, simultaneous inference allows there to be feedback between the cheap and expensive model emulators. Moreover, \cite{jck_ml} analytically integrates out the regression parameters for the emulator mean function. The resulting predictive processes using \cite{cite:13} and our adaptation of \cite{jck_ml}'s approaches are similar in structure, yet differ in how they estimate their emulation parameters.

As in \cite{jck_ml}, we aim to construct a high-fidelity emulator that accurately represents the true computer model. We follow their general emulation approach and notation. Here we fit an emulator separately to each column of $\bY^R = ( \bY^R_1 ,..., \bY^R_{J_y} )$, the principal components of the computer model runs. To simplify notation we denote column $j$ of $\bY^R$ as $\bt=\bY^R_j$. We also denote $\bt^E= \bY^{RE}_j = ( Y^R_{1,j} ,..., Y^R_{p^E,j} ) ^T$ as the entries corresponding to the expensive model runs at design points $\{\btheta_1^E,..., \btheta^E_{p^E}\}$, and $\bt^C= \bY^{RC}_j = ( Y^R_{p^E+1,j} ,..., Y^R_{p^E+p^C,j} ) ^T$ as the entries corresponding to the cheap model runs at design points $\{\btheta_1^C,..., \btheta^C_{p^C}\}$.

The probability model proposed by \cite{jck_ml} fit to $\bt$ is as follows:
\begin{gather}\label{EQ: JCK_MVN}
 \begin{bmatrix} \bt^C \\ \bt^E \end{bmatrix} |\bOmega
 \sim
 \mcN \Bigg( 
  \begin{bmatrix}
   h(\btheta^C) \bbe^C \\
   h(\btheta^E) (\rho \bbe^C + \bbe^E)
   \end{bmatrix},
   \begin{bmatrix}
   Cov(\bt^C| \bOmega) &
   Cov(\bt^C,\bt^E|\bOmega) \\
   Cov(\bt^E,\bt^C|\bOmega) &
   Cov(\bt^E| \bOmega)
   \end{bmatrix}
   \Bigg)
\end{gather}
\noindent 
where $\btheta^C = (\btheta_1^E,..., \btheta^E_{p^E})^T$ and $\btheta^E= (\btheta_1^C,..., \btheta^C_{p^C})^T$ are the matrices of design points for the cheap and expensive model runs. $h()$ is the mean function of the matrix of parameter settings; here $h(\btheta^E)= (\bOne, \btheta^E)$ and $h(\btheta^C)= (\bOne, \btheta^C)$. $\bbe^C$ and $\bbe^E$ are the coefficient vectors for $h(\btheta^C)$ and $h(\btheta^E)$. $\rho$ is a cross-correlation parameter that describes the amount of information that is borrowed from the lower resolution runs in emulating the higher-resolution runs. $\bOmega= \{\rho, \bbe^C, \bbe^E, \lambda^2_C, \lambda^2_E, \phi^C, \phi^E, \sigma^2_C, \sigma^2_E\}$ is the set of all emulator parameters, where $\{\lambda^2_C,\sigma^2_C,\phi^C,\lambda^2_E,\sigma^2_E,\phi^E\}$ are covariance parameters for the cheap and expensive model runs.

Following \cite{jck_ml}, we define the covariance of $\bt^C$ and $\bt^E$ and the cross-covariance between $\bt^C$ and $\bt^E$ using the squared exponential covariance function.

We calculate these quantities by:
\begin{align*}
    C_C(\btheta^C_i,\btheta^C_j|\bOmega)=& \sigma^2_C exp{-(\btheta_i^C-\btheta_j^C)'\bD_C^{-1}(\btheta_i^C-\btheta_j^C)} + \lambda_C^2 I(\btheta_i^C=\btheta_j^C),\\
    C_E(\btheta^E_i,\btheta^E_j|\bOmega)=& \rho^2 \sigma^2_C exp{-(\btheta_i^E-\btheta_j^E)'\bD_C^{-1}(\btheta_i^E-\btheta_j^E)} +\\
    & \sigma^2_E exp{-(\btheta_i^E-\btheta_j^E)'\bD_E^{-1}(\btheta_i^E-\btheta_j^E)}
    + \lambda_E^2 I(\btheta_i^E=\btheta_j^E),\\
    \textnormal{and} \\
    C_{CE}(\btheta^C_i,\btheta^E_j|\bOmega)
    =& \rho \sigma^2_C exp{-(\btheta_i^C-\btheta_j^E)'\bD_C^{-1}(\btheta_i^C-\btheta_j^E)}.
\end{align*}
\noindent $\bD_C$ is a $k \times k$ diagonal matrix with elements $\{\phi_1^C,..., \phi_k^C\}$, where $k= dim(\btheta^C_i)$. $\bD_E$ is the analogue for the expensive model output. Although our computer model is deterministic and no nugget is needed, we include a nugget term. Many researchers advocate for this practice: one issue with zero-nugget emulators is numerical instability \citep{ra_acb_efw1994,neal_1997}. In addition, \cite{rbg_hkhl2012} show that when the training data is sparse or common violations of standard assumptions occur, including a non-zero nugget yields better emulator coverage and predictive accuracy. The GPMSA (Gaussian Process Models for Simulation Analysis) package \citep{Gattiker2016} also estimates a small nugget term when emulating deterministic computer models.

By assigning the priors $(\bbe^C , \bbe^E)^{T}\sim \mcN(\bb,\bB)$, where $\bB$ is a block diagonal matrix with $\bB_C$ at the top left and $\bB_E$ at the bottom right, we can integrate out $\bbe^C$ and $\bbe^E$ \citep{jck_ml}. We define $\bOmega_{-\beta}$ to be the set of emulator parameters without $\bbe^C$ and $\bbe^E$. The distribution of $\bt$ conditional on $\bOmega_{-\beta}$ is:
\begin{align}
    \bt|\bOmega_{-\beta} \sim \mcN(\bH \bb,Var(\bt|\bOmega_{-\beta})+ \bH \bB \bH'),
\end{align}
where $\bH$ is the block matrix:
\begin{gather}
   \bH= \begin{bmatrix}    
   h(\btheta^C) &
   \bzero \\
   \rho h(\btheta^E) &
   h(\btheta^E)
   \end{bmatrix}.
\end{gather}

\newpage
\noindent Following \cite{jck_ml} we adopt the following priors for $\bOmega_{-\beta}$:

\vspace{-0.5in}
\begin{multicols}{2}
\begin{align*}
    \lambda^2_C &\sim \textnormal{Inverse Gamma}(\alpha_{\lambda_C},\beta_{\lambda_C}) \\
    \phi^C &\sim \textnormal{Gamma}(\alpha_{\phi^C},\beta_{\phi^C}) \\
    \sigma^2_C &\sim \textnormal{Inverse Gamma}(\alpha_{\sigma_C},\beta_{\sigma_C})\\
    \rho &\sim N(\mu_\rho,\sigma^2_\rho)
    \end{align*}
    
    \begin{align*}
    \lambda^2_E &\sim \textnormal{Inverse Gamma}(\alpha_{\lambda_E},\beta_{\lambda_E}) \\
    \phi^E &\sim \textnormal{Gamma}(\alpha_{\phi^E},\beta_{\phi^E}) \\
    \sigma^2_E &\sim \textnormal{Inverse Gamma}(\alpha_{\sigma_E},\beta_{\sigma_E})
\end{align*}
\end{multicols}
\vspace{-0.25in}
\noindent We obtain the maximum a posteriori point estimates of the emulator parameters $\bOmega_{-\beta}$ using the optimizing function in RStan \citep{stan}, which employs the Limited-memory Broyden–Fletcher–Goldfarb–Shanno (L-BFGS) algorithm \citep{byrd1995limited}. If the estimated value of $\rho$ is zero, this means there is no linear relation between the means of  $\bt^C$ and $\bt^E$.
The multiresolution Gaussian process emulator is designed to predict the value of $t^E$ at untried design point $\btheta_0$ and is defined by its mean and covariance:
\begin{align}\label{EQ:emulatorMean}
\begin{split}
 E[t^E(\btheta_0)&|,\bt^E,\btheta^E,\bt^C,\btheta^C,\bOmega_{-\beta}] = \\ 
   &\begin{bmatrix}    
   \rho h(\btheta_0) & h(\btheta_0)
   \end{bmatrix}
   \bb + Cov(t^E(\btheta_0),\bY) (Var(\bt|\bOmega_{-\beta}) + \bH\bB \bH')^{-1} (\bt- \bH \bb)
\end{split}
\end{align}

\begin{align}\label{EQ:emulatorCov}
\begin{split}
 Cov[t^E(\btheta_0)&|,\bt^E,\btheta^E,\bt^C,\btheta^C,\bOmega_{-\beta}] = \\ 
    &\rho^2 C_C(\btheta_0,\btheta_0) + C_E(\btheta_0,\btheta_0) + h(\btheta_0) (\rho^2 \bB_C + \bB_E) h(\btheta_0)' \\
    &- Cov(t^E(\btheta_0),\bt) (Var(\bt|\bOmega_{-\beta}) + \bH\bB \bH')^{-1} Cov(t^E(\btheta_0),\bt)' + \lambda^2_E
\end{split}
\end{align}

\section{4. Application to the LISFLOOD-FP flood hazard model}\label{Sec:AppLISFLOOD-FP} 
In this section, we provide implementation details and discuss the calibration results for the LISFLOOD-FP flood hazard model using the multiresolution approach (MR). 
In addition, we conduct a comparative analysis against a single-resolution approach using just the high resolution model runs (HR).

We evaluate our approach's accuracy in inferring $\btheta^*= (n_{ch}^*,RWE^*)$ for two different values of $\btheta^*$, denoted as $\btheta^{*1}$ and $\btheta^{*2}$ . In both cases, we generate a single expensive model run from LISFLOOD-FP at a selected $\btheta^*$. Next, we contaminate this model run with independent and identically distributed Gaussian random noise at locations where nonzero flood heights were observed. The values of flood heights are truncated below at zero to ensure physically realistic observations. We set the standard deviation of the Gaussian random noise to be 3cm due to concerns about introducing too much positive bias to the amount of flooding observed. 

In the first case, $\btheta^{*1}= (n_{ch}^{*1}=0.0305, RWE^{*1}= 1)$. $RWE^{*1}$ is chosen to reflect that the estimate of river width has no error. The chosen $n_{ch}^{*1}$ is based on field observations, topography, expert judgment, and photography \citep{snydercounty_2007}. For the second case, we consider a parameter set $\btheta^{*2}= (n_{ch}^{*2}=0.0249, RWE^{*2}= 1.0452)$ where the true parameters lie near the edge of the parameter space. When only a small number of expensive model runs are available, there may not be enough design points at a critical region (i.e. near the edge of the parameter space) for accurate inference. Hence, we aim to improve inference by running the cheaper model runs at carefully placed design points, particularly near the boundaries. 
To study this case, we consider the scenario where no expensive model runs are available at the edge of the parameter space, precisely where $\btheta^{*2}$ lies. Consequently, only the cheap model runs provide information pertaining to the edges of the parameter space. To set up this scenario, we exclude the expensive design points where $RWE$ is in the top 5\% of its plausible range ($RWE \in (1.045,1.05)$) or $n_{ch}$ is in the bottom 5\% of its plausible range ($n_{ch} \in (0.02,0.028)$). We refer to this scenario as the `edge case.'


In this study, we examine two cases to understand whether information from cheap model runs can improve calibration when different numbers of expensive model runs are available. We consider the following combinations of numbers of expensive ($N^E$) and cheap ($N^C$) model runs: (a) 50 expensive and 200 cheap and (b) 100 expensive and 400 cheap. In both combinations all the design points for the expensive model runs, i.e. expensive design points, are selected using a maximin Latin square design. We run the cheap model at the expensive design points and either 150 or 300 additional design points selected using a maximin Latin square with the `lhs' R package \citep{lhs}.

We set the following prior distributions for the multiresolution emulator parameters:
\newpage
\begin{multicols}{2}
\begin{align*}
    \lambda^2_C &\sim \textnormal{Inverse Gamma}(2,2)\\
    \theta^C &\sim \textnormal{Gamma}(2,2) \\
    \sigma^2_C &\sim \textnormal{Inverse Gamma}(2,2) \\
    \bbe &\sim \mcN(\bzero,\bI)
    \end{align*}
    
    \begin{align*}
    \lambda^2_E &\sim \textnormal{Inverse Gamma}(2,2)\\
    \theta^E &\sim \textnormal{Gamma}(2,2) \\
    \sigma^2_E &\sim \textnormal{Inverse Gamma}(2,2) \\
    \rho &\sim \mcN(1,\frac{1}{3})
    \end{align*}
\end{multicols}
\vspace{-0.1in}

\noindent where $\bbe= \begin{bmatrix} \bbe^C & \bbe^E \end{bmatrix}'$. Following \citet{jck_ml}, we place an informative prior $\mcN(1,\frac{1}{3})$ on $\rho$, which assumes the cheap model runs should be informative for the expensive model runs. Upon fitting the multiresolution emulator, we perform calibration using the simulated observation at $\btheta^{*1}$ and at $\btheta^{*2}$. 
We set the priors to be $n_{ch} \sim \mathcal{U}(0.02,0.1)$ and  $RWE \sim \mathcal{U}(0.95,1.05)$ to reflect no additional knowledge about the parameter values aside from the expert-determined plausible range.



\subsection{ 4.1 Comparisons to standard calibration approach}
We conduct a comparative study between our multiresolution (MR) emulation-calibration approach and traditional single-resolution (HR) emulation-calibration only using the expensive high resolution model runs. $N^E$e$N^C$c indicates the number of expensive ($N^E$) and cheap ($N^C$) runs used in the MR approach; in the corresponding HR approach, only $N^E$ expensive runs are used. In the `edge case' study, these are the numbers before removing the model runs at edge parameter settings. We first compare how well the MR and HR emulators approximate the expensive computer model runs. We then compare how accurately the MR and HR emulation-calibration approaches infer the the parameter setting for each simulated observation. Finally, we compare the calibrated predictions to the simulated observations.

\subsubsection{4.1.1 Emulation}
We compare these emulators' performances both in the context of a 10-fold cross-validation study and in the context of an edge case study. In the 10-fold cross validation study 10\% of the expensive model runs are held out at random, the emulators are trained on the rest of the model runs, and their predictive abilities are compared on the held out expensive model runs. In the the edge case study, we hold out the expensive model runs at design points where $RWE \in (1.045,1.05)$ or $n_{ch} \in (0.02,0.028)$ so that the only model runs at edge design points are cheap. We then train the emulator on the remaining model runs, and compare the predictive abilities of the MR and HR emulators at the held out design points. We perform both the cross-validation study and edge case study for the combinations of expensive and cheap model runs previously listed. 

At the original scale of the model runs, we compare the performances of the emulators in terms of the root mean-squared error (RMSE). We define $D_{MR-HR}= RMSE_{MR} - RMSE_{HR}$, i.e. the difference in RMSE between the MR emulator and the HR emulator at the same design point. A negative $D_{MR-HR}$ indicates that the MR approach had smaller RMSE, and a positive $D_{MR-HR}$ indicates that the HR approach had smaller RMSE. In the cross-validation study (Table \ref{Table: CVRMSE}), the summary statistics for $D_{MR-HR}$ indicate that the MR emulator performs better in terms of RMSE in the 50e200c combination: here the median and mean $D_{MR-HR}$ are -0.03 and -0.033, while for the 100e400c combination the median and mean of $D_{MR-HR}$ are both 0.015. These summary statistics reveal a small difference in performance of the two emulators in terms of RMSE for either combination, as the unit of measurement is in m and flood heights projected by the computer model range from 0 to over 12.5 m.

\begin{table}[h]
\caption{Cross validation: difference in RMSE between MR and HR ($D_{MR-HR}$) (m)}
\centering
\begin{tabular}{rllll}
  \hline
 & Q1  & Median & Mean & Q3 \\ 
  \hline
  50e200c  & -0.180 & -0.030 & -0.033 & 0.047 \\ 
  100e400c & 0.007 & 0.015 & 0.015 & 0.022 \\
   \hline
\end{tabular}

\label{Table: CVRMSE}
\end{table}

We also consider the the values of $D_{MR-HR}$ for the edge case study (Table \ref{Table: CVRMSE Edge}). For 50e200c but not 100e400c the MR emulator outperforms the HR emulator in terms of RMSE at the edge design points. For 50e200c, the difference in performances (median $D_{MR-HR}$ = -0.533, mean $D_{MR-HR}$ = -0.736) is much larger than for 100e400c (median $D_{MR-HR}$ = 0.035, mean $D_{MR-HR}$ = 0.054). In both the cross-validation and edge case studies, the values of the raw RMSEs reveal that the 100e400c HR and MR emulator have much greater predictive accuracy than the the 50e200c HR and MR emulators. 

\begin{table}[h]
\caption{Edge cases: difference in RMSE between MR and HR ($D_{MR-HR}$) (m)}
\centering
\begin{tabular}{rllll}
  \hline
 & Q1  & Median & Mean & Q3 \\ 
  \hline
  50e200c  & -1.426 & -0.533 & -0.736  & 0.165 \\ 
  100e400c & 0.012 & 0.035 & 0.054 & 0.109 \\
   \hline
\end{tabular}
\label{Table: CVRMSE Edge}
\end{table}

To sidestep the computational costs of evaluating other aspects of the emulator's performance at the original scale, we consider the uncorrelated standardized prediction errors (USPEs) \citep{bastos_ohagan2009} for the leading principal components for each emulation approach, as done by \cite{chang2014}. We define leading principal components as defining at least 75\% of the variation. We plot the uncorrelated standardized prediction errors against the theoretical quantiles of the appropriate T distribution (degrees of freedom = number of points in test set - number of parameters estimated in the mean) \citep{bastos_ohagan2009}, the emulator prediction, channel roughness, and river width error to compare MR versus HR emulator fit. We provide these plots for the cross-validation and edge case study in the supplement.

First we consider these plots in the cross validation study. For 50e200c, the $|USPE|$s tend to be smaller for the MR approach (Q1= -0.148, Q3=0.219, |Max|= 8.29) than for the HR approach (Q1= -0.614, Q3=0.462, |Max|= 15.2). Considering the QQ plots and confidence bounds, both approaches exhibit good fit in the middle and heavy tails. However the MR approach's lower tail is closest to appropriate thickness. 
In the plots of the USPEs versus $n_{ch}$, we see that for both the MR and HR approaches there is no general pattern but slightly increased spread in some areas. In the plots of the USPEs versus $RWE$, there are ranges of $RWE$ with increased variability in the USPEs for both approaches: for the MR approach this is near the upper boundary and for the HR approach this is slightly above the lower boundary. For the MR approach, the plot of the predicted values versus the USPEs gives no indication of heteroscedasticity. For the HR appraoach, this plot shows slight increased spread for medium-high predictions. Overall, the USPE plots indicate better performance of the MR emulator. 

For 100e400c, the $|USPE|$s again tend to be smaller for the MR approach (Q1= -0.075, Q3= 0.111, |Max|= 1.44) than for the HR approach (Q1= -0.570, Q3= 0.575, |Max|= 9.95). Considering the QQ plots and confidence bounds, for the MR emulator, the fit is good in the middle but unsatisfactory in the tails. For the HR emulator, the same is true except the lack of fit in the upper tail is more pronounced. Considering all plots of the USPEs versus $RWE$, we see that in both approaches there is no general pattern. In the plots of the USPEs versus $n_{ch}$, we see that for the neither approach there is a strong pattern, but for the HR approach there may be slightly decreased spread at the highest $n_{ch}$ values. Concerning the plots of USPEs versus the predicted values, for the MR approach there may be slightly increased spread at the highest and lowest predicted values, while for the HR approach there may be slightly increased spread at the middle predicted values. However, neither plot shows a strong pattern. Based on the USPEs, we show preference to the MR emulator.
In the supplement, we provide a detailed examination of the USPEs for the edge case model runs for the 50e200c and 100e400c settings. Similar to the non-edge cases, the $|USPE|$s tend to be smaller for the MR approach than the HR approach; hence, the MR emulator is preferred over the HR counterpart. 



\subsubsection{4.1.2 Calibration}

\begin{figure}
     \centering
     \begin{subfigure}[b]{0.48\textwidth}
         \centering
         \includegraphics[width=\textwidth]{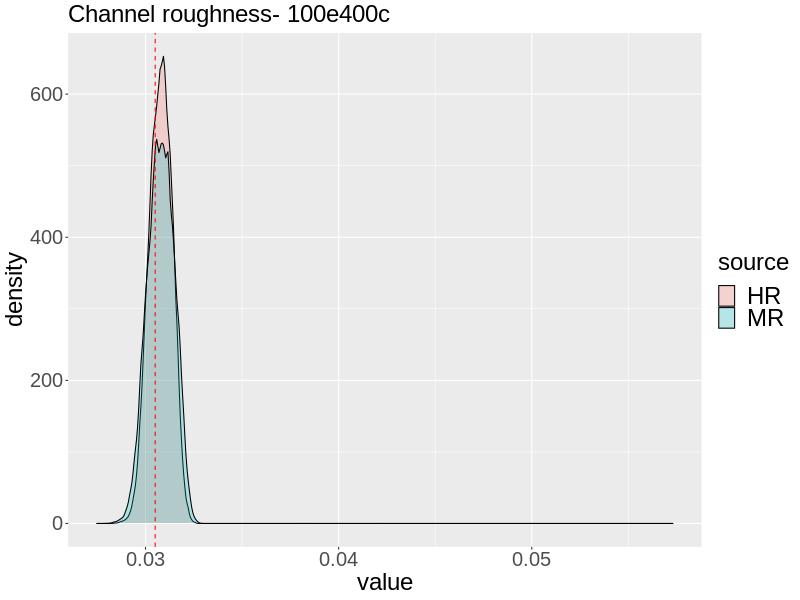}
         \caption{$n_{ch}=0.0305$}
         \label{fig: Ch Posterior}
     \end{subfigure}
     \centering
     \begin{subfigure}[b]{0.48\textwidth}
         \centering
         \includegraphics[width=\textwidth]{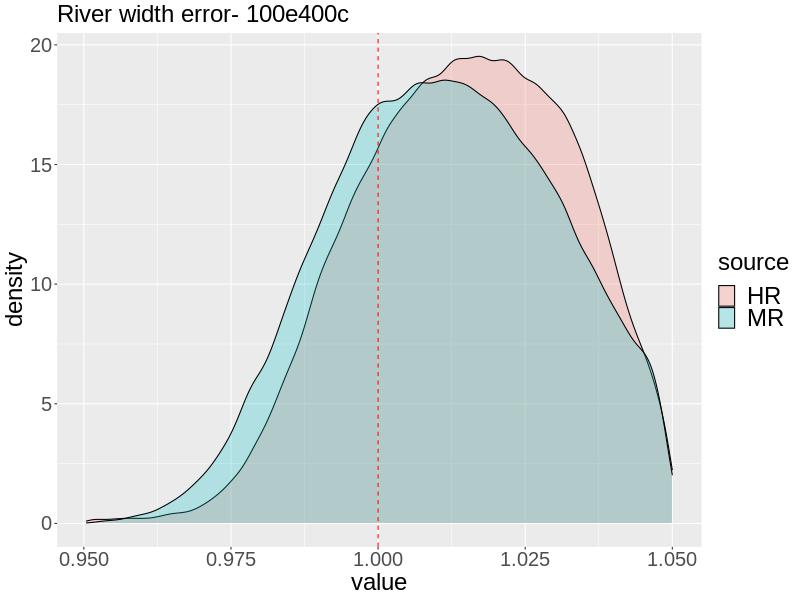}
         \caption{$RWE=1$}
         \label{fig: RWE Posterior}
     \end{subfigure}
     \begin{subfigure}[b]{0.48\textwidth}
         \centering
         \includegraphics[width=\textwidth]{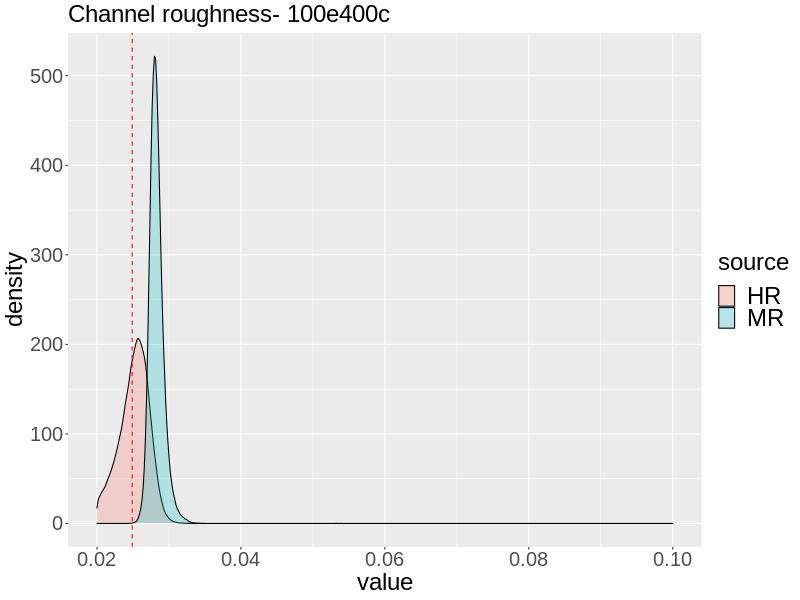}
         \caption{$n_{ch}=0.0249$}
         \label{fig: Ch Posterior lohi}
     \end{subfigure}
     \centering
     \begin{subfigure}[b]{0.48\textwidth}
         \centering
         \includegraphics[width=\textwidth]{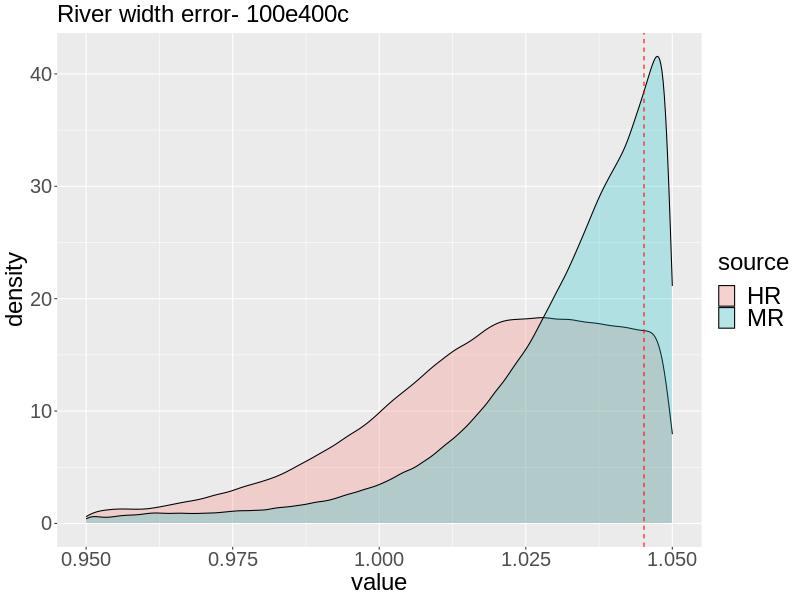}
         \caption{$RWE=1.0452$}
         \label{fig: RWE Posterior lohi}
     \end{subfigure}
     \caption{Posterior densities from each calibration approach generated either using the MR (blue) or HR (pink) for simulated observation 1 (top) or simulated observation 2 (bottom)}
    \label{Fig: 100e400cPosteriors}
\end{figure}

Next, we compare the calibration results using the MR versus the HR approach. For both approaches, we ran the MCMC algorithm for 300,000 iterations to ensure satisfactory effective sample size ($\geq 3500$). We display the calibration results using the 100e400c combination of model runs in the main text (Figure \ref{Fig: 100e400cPosteriors}) as these had greatest predictive accuracy in terms of RMSE for the cross validation and edge case studies. We provide the calibration results using other combinations of model runs in the supplement. 

For the 100e400c combination when $\btheta^{*1}=(n_{ch}^{*1}=0.0305, RWE^{*1}= 1)$, $n_{ch}^{*1}$ is inferred nearly accurately by both approaches, but slightly overestimated by both. The uncertainty of the MR approach is slightly greater, and the HR approach places slightly more posterior density at $n_{ch}^{*1}=0.0305$. Both approaches yield a posterior mean of 0.0308. For both the MR and HR approaches, the posterior for $RWE^{*1}$ is fairly deconcentrated and contains the truth, but the MR approach is slightly more accurate, having its center closer to and more posterior density at $RWE^{*1}=1$. This is reflected in the posterior means from the MR and HR approaches: 1.011 and 1.015, respectively.

For $\btheta^{*2}=(n_{ch}^{*2}=0.0249,RWE^{*2}=1.0452)$ with 100e400c, the HR approach estimates $n_{ch}^{*2}$ very accurately with posterior mean 0.0252. The MR slightly overestimates $n_{ch}^{*2}$, giving a posterior mean of 0.0283, with too sharp of a posterior such that almost no posterior density is allocated to $n_{ch}^{*2}=0.0249$. The MR approach estimates $RWE^{*2}=1.0452$ fairly accurately with much of the posterior density being allocated there, but with the posterior being skewed left, the posterior mean is 1.031. The HR approach yields a wider posterior density that contains $RWE^{*2}=1.0452$ but is not as informative, giving a posterior mean of 1.019. 

We also obtain projections from the expensive model using thinned posterior parameter samples for each $\btheta^*$ from each calibration approach. These projections come from using 100e400c since this combination generally produced the most reasonable emulation and calibration results across approaches. We thin the MCMC samples because obtaining a model projection at all sampled parameter values would be computationally infeasible. 
We selected 100 thinned samples to be consistent with the number of expensive model runs afforded for training the emulators. We run LISFLOOD-FP at the parameter values of each selected step. For each calibration method, we calculate the mean of the projections from the thinned posterior and call this the calibrated projection. We compare the mean projections to the expensive projections for each $\btheta^*$. 

For $\btheta^{*1}$ and $\btheta^{*2}$ we compare the calibrated projections for each method to the simulated observation using root mean-squared error, fit, percent bias, and correctness. We calculate fit and correctness by $\textnormal{fit} = (A_{rm})/(A_r + A_m - A_{rm})$ and $\textnormal{correctness} = (A_{rm})/(A_r)$, where $A_r$ is the observed flooded area, $A_m$ is the projected flooded area, and $A_{rm}$ is the area both observed and projected to be flooded \citep{Rajib_etal_2020}.
$\textnormal{Percent bias}= 100\times(\sum_{j=1}^{N_L} (P_j-Z_j))/( \sum_{j=1}^{N_L} Z_j)$,
where $Z_j$ is the simulated observation at location $j$ and $N_L$ is the number of locations \citep{PBiasSource}. $P_{j}$ is the mean projection at location $j$ resulting from the given calibration method. 

\begin{figure}
     \centering
     \begin{subfigure}[b]{0.48\textwidth}
         \centering
         \includegraphics[width=\textwidth]{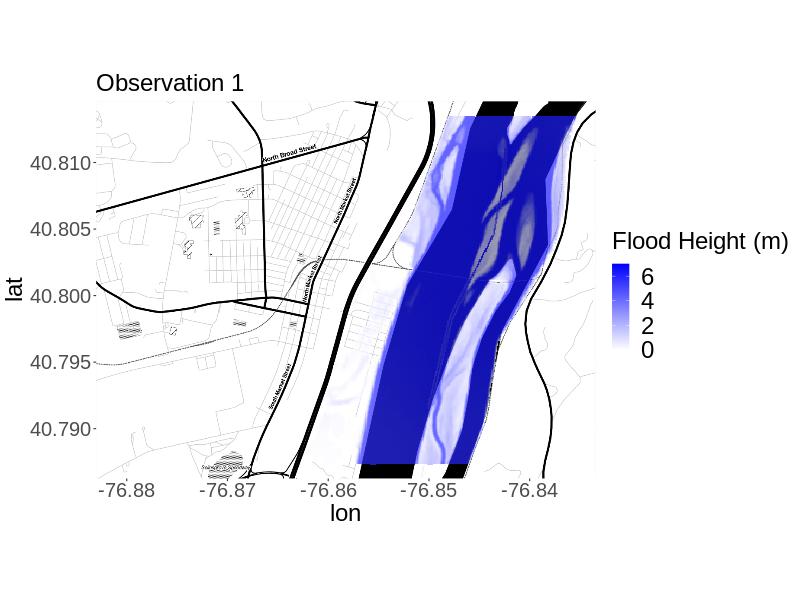}
         \caption{Observed flood heights for $\btheta^{*1}$}
         \label{fig: Observed flood heights 1}
     \end{subfigure}
     \hfill
     \begin{subfigure}[b]{0.48\textwidth}
         \centering
         \includegraphics[width=\textwidth]{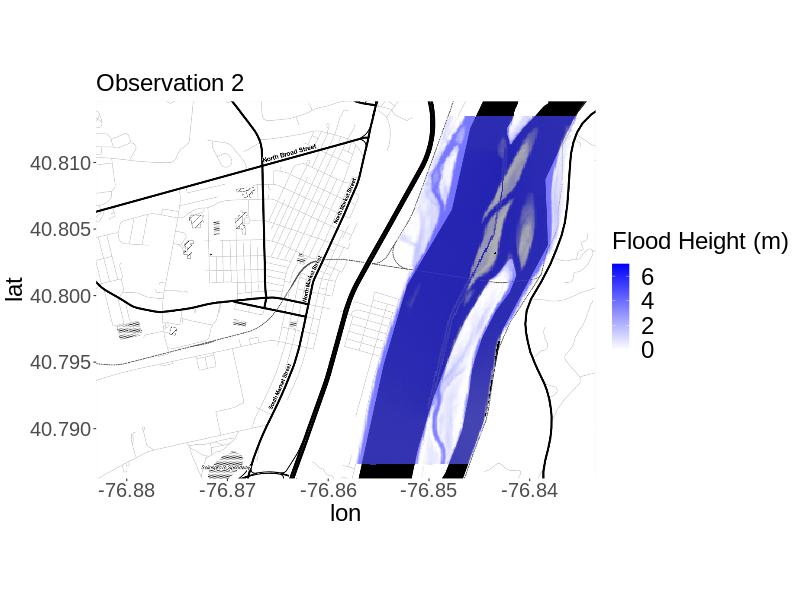}
         \caption{Observed flood heights for $\btheta^{*2}$}
         \label{fig: Observed flood heights 2}
     \end{subfigure}
     \begin{subfigure}[b]{0.48\textwidth}
         \centering
         \includegraphics[width=\textwidth]{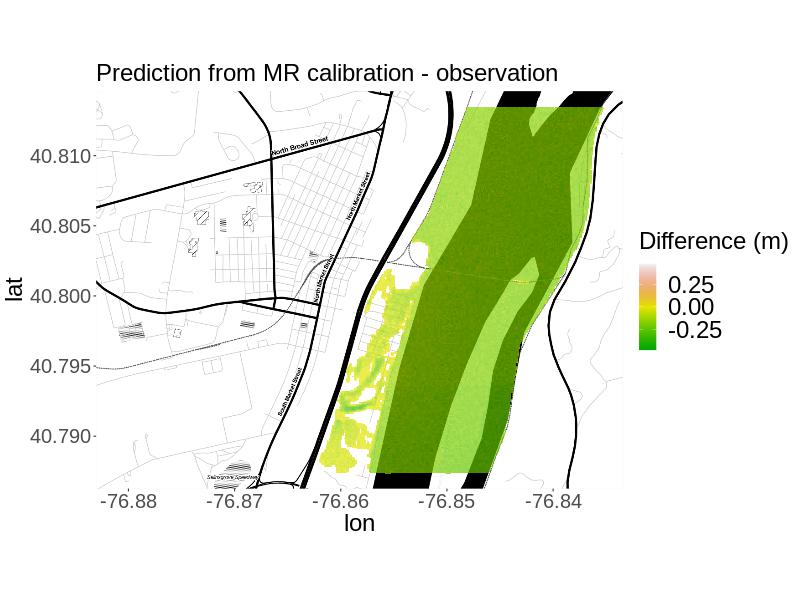}
         \caption{MR residuals for $\btheta^{*1}$}
         \label{fig:HomMR residuals 1}
     \end{subfigure}
     \hfill
     \begin{subfigure}[b]{0.48\textwidth}
         \centering
         \includegraphics[width=\textwidth]{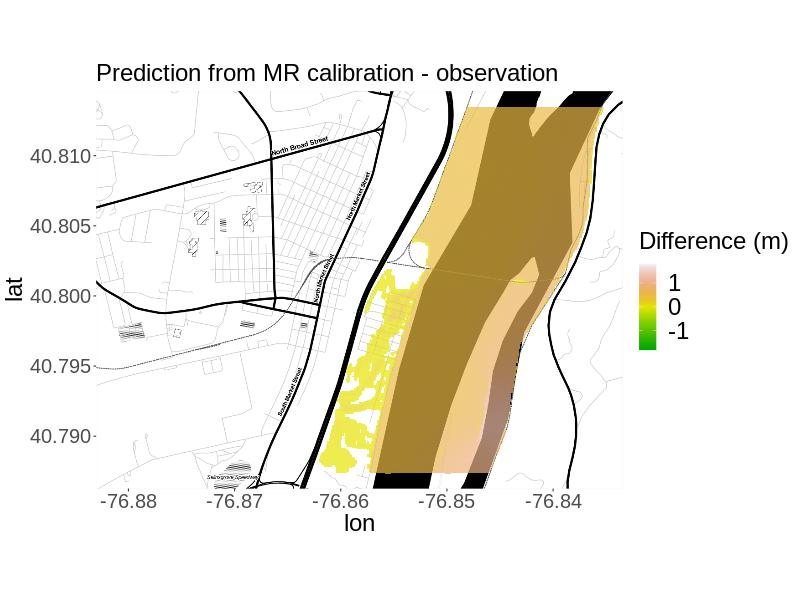}
         \caption{MR residuals for $\btheta^{*2}$}
         \label{fig:HomMR residuals 2}
     \end{subfigure}
     \begin{subfigure}[b]{0.48\textwidth}
         \centering
         \includegraphics[width=\textwidth]{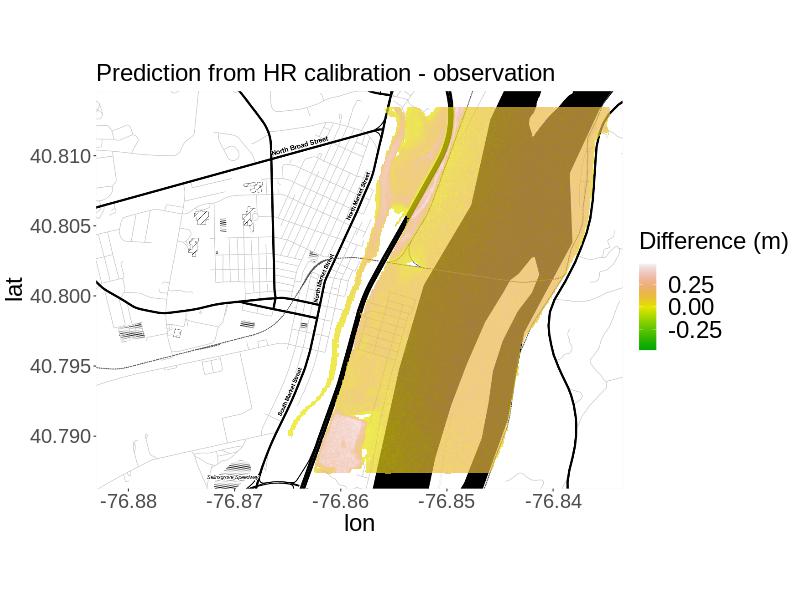}
         \caption{HR residuals for $\btheta^{*1}$}
         \label{fig:HomGP residuals 1}
     \end{subfigure}
     \hfill
     \begin{subfigure}[b]{0.48\textwidth}
         \centering
         \includegraphics[width=\textwidth]{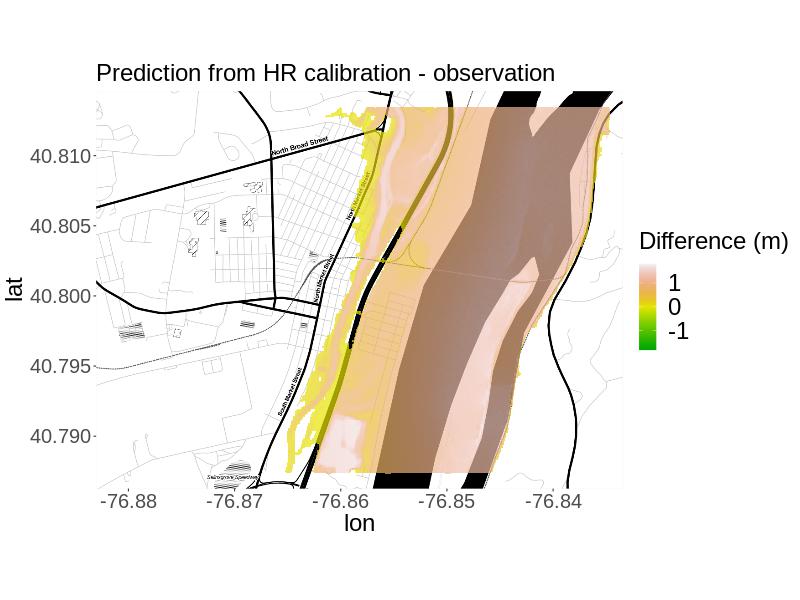}
         \caption{HR residuals for $\btheta^{*2}$}
         \label{fig:HomGP residuals 2}
     \end{subfigure}
     \caption{$\btheta^{*1}=(n_{ch}=0.0305,RWE=1)$: Simulated observation (a), difference between calibrated projection from MR (c), HR (e) and the observation. $\btheta^{*2}= n_{ch}=(0.0249,RWE=1.0452)$: Simulated observation (b), difference between calibrated projection from MR (d), HR (f) and the observation.}
    \label{Fig: ObservedVSResidualsMap}
\end{figure}

The plotted results (Figure \ref{Fig: ObservedVSResidualsMap}) and the metrics (Table \ref{Table: ED_F_C_MeanPredictions_Real}) show that for $\btheta^{*1}= (n_{ch}= 0.0305,RWE=1)$, both calibration approaches yield projections very similar to the simulated observation. The RMSE for the MR approach is slightly higher, but by less than 5 cm. Both approaches have nearly identical percent bias but the MR approach's bias (-3.73\%) is negative and the HR approach's (3.72\%) is positive, as reflected by the plots of the residuals in space. However, the fit indicates that the the MR calibrated projection was notably closer to the simulated observation in terms of predicting the same areas as flooded versus not flooded (98.5\% versus 78.6\% overlap). Correctness was equal to one for both approaches, so the calibrated projections from both approaches predicted all flooded areas to be flooded. In summary, if the primary interest is in projecting flooding in the right areas, the MR approach is preferable, but if the primary interest is in predicting high enough flood heights, the HR approach is preferable.

For $\btheta^{*2}= (n_{ch}=0.0249,RWE=1.0452)$, the plotted results (Figure \ref{Fig: ObservedVSResidualsMap}) and the metrics (Table \ref{Table: ED_F_C_MeanPredictions_Real}) indicate that the MR approach yields a calibrated projection most similar to the simulated observation considering the plotted results and metrics. Both approaches yield calibrated projections with more flooding than in the observation, as reflected by the percent bias and in the plots of the residuals. Specifically, the RMSE is smaller, the absolute value of the percent bias is smaller, and the overlap in projected flooded areas was much larger. Based on calibrated projections for this simulated truth, we would recommend the MR approach. 

\begin{table}[h]
\caption{Comparison of mean projection to observation 1 ($n_{ch}= 0.0305,RWE=1$)}
\centering
\begin{tabular}{rllll}
  \hline
  & MR & HR\\ 
  \hline
  RMSE & 0.0908 & \textbf{0.0860} \\
  Percent Bias & -3.73 &  \textbf{3.72} \\
  Fit & \textbf{0.985} & 0.786 \\
  \hline
\end{tabular}
\label{Table: ED_F_C_MeanPredictions_Real}
\end{table}

\begin{table}[h]
\caption{Comparison of mean projection to observation 2 ($n_{ch}=0.0249,RWE=1.0452$)}
\centering
\begin{tabular}{rllll}
  \hline
  & \textbf{MR} & HR\\ 
  \hline
  RMSE & \textbf{0.298} & 0.658 \\
  Percent Bias & \textbf{14.7} & 39.3 \\
  Fit & \textbf{0.879} & 0.625 \\
  \hline
\end{tabular}
\label{Table: ED_F_C_MeanPredictions_Edge}
\end{table}

\subsubsection*{Computational Costs}
With regards to computational cost, our approach is sensible for cases where the difference in model run times between the expensive and cheap model runs is larger. We compute the amount of time each calibration approach needed to reach 300,000 steps using the variable-at-a-time random walk Metropolis-Hastings algorithm for the 100e400c combination (Table \ref{Table: CalTimes}). Generally, the multiresolution calibration methods are 5.79-6.31 times slower, likely due to the larger matrix operations associated with the additional cheap model runs and a more complicated emulator structure. 

To obtain the LISFLOOD-FP model runs, obtaining 100 expensive model runs sequentially takes about 10.2 hours. Obtaining 400 cheap model runs sequentially takes 1.78 hours. Some of these model runs may be obtained in parallel depending on the computing system available, however parallelizing the expensive model runs makes model failure more likely. We also computed the amount of time needed for interpolating the cheap model runs and computing principal components and emulation for each approach. Interpolation took 7.81 minutes for the 100e400c combinations. Computing principal components took no more than 30 seconds for any approach. Fitting the emulators to all principal components sequentially takes 1.25 to 6.05 minutes for the MR approaches, and 5.86 to 9.36 seconds for the HR approaches depending on the combination of model runs.

\begin{table}[h]
\caption{Calibration times in hours}
\centering
\begin{tabular}{rllll}
  \hline
  Observation 1 ($n_{ch}=0.0305,RWE=1$)\\ 
  \hline
  & MR & HR\\ 
  \hline
  100e400c & 23.40 & 4.04 \\
  \hline
  Observation 2 ($n_{ch}=0.0249,RWE=1.0452$) \\
  \hline
  & MR & HR\\ 
  \hline
  100e400c & 22.31 & 3.54 \\
\end{tabular}
\label{Table: CalTimes}
\end{table}

\section{5. Discussion}\label{Sec:Discussion}

\subsection{5.1 Caveats}

Due to the limited availability of flood height measurements and satellite imagery for historical flooding in Selinsgrove, we test our approach using two simulated observations. Simulated observations may not completely capture the true observations for the flood events they aim to mimic. In addition, because we used a simulated observation with independent and identically distributed Gaussian random noise, there was no need to estimate a model-observation discrepancy, $\delta$, which tends to be important and challenging to estimate when calibrating a model using real observations. In future research, we plan on either switching our focus to a river with more observations available or making use of citizen science in Selinsgrove to obtain real observations. We also plan on varying the ratio of expensive to cheap computer model model runs in future work to explore how the emulation and calibration results change. Perhaps there is a model-specific `best' ratio for calibration. 

While we only explore one edge case scenario, it is possible that the performance of the multiresolution calibration approach may change at different edges of the parameter space. In addition, our results are specific to our adapted version of \cite{jck_ml}'s multiresolution emulator. A comparison with the emulation approach from \citet{cite:13} may be a useful study. With respect to inference using calibrated projections, it is important to note that results will vary depending on the thinning process. 

Finally, an important issue to note is that parameter interpretation can often change with changes to model resolution. Hence, there are some challenges with interpreting calibration results that use model runs from different resolutions. We have not attempted to tackle this difficult question as we view our work here as merely a first step toward understanding the impact of calibration based on multiresolution model runs. 

\subsection{5.2 Summary}

We have developed a new calibration approach that allows researchers to combine information from model runs across different model resolutions. To our knowledge, this is the first such approach and likely to be useful as the ability to use model runs at different resolutions is potentially valuable in many research and engineering problems. On the other hand, we believe our study has barely scratched the surface of many interesting challenges and questions arising from the study of models at different resolutions. 

We studied the application of our method to the LISFLOOD-FP flood hazard model and find that the results vary. We employ a multiresolution approach that uses low resolution model runs in addition to high resolution model runs.
Using the appropriate combination of high resolution and low resolution model runs, we find that the multiresolution calibration approach more accurately infers river width error both at the edge and in the middle of the plausible range of values. However, we find that our approach has little to offer in inferring channel roughness over calibration with just the high resolution runs. 
Studies have found that flood models are generally very sensitive to channel roughness while river width does not seem to be a sensitive parameter \citep{pappenberger_2008,savage_etal2016,alipour_etal2022}. This means that river width error may require more model runs for accurate inference than channel roughness, and based on our results with the appropriate number of model runs, the multiresolution calibration approach could aid in meeting that requirement.
Within the context of LISFLOOD-FP projections, the multiresolution approach provide improved projections of flooded locations over the single-resolution approach

Studying models at different resolutions is a very complicated problem as there are a large number of variables that can impact the conclusions:
\begin{enumerate}
    \item The choice of the number of high and low resolution model runs. There are complex tradeoffs between parameter interpretation across resolutions, computational costs for running the model at different resolutions, as well as the cost of running different calibration approaches.
    \item The design used for the parameter sets used in each case. The design for the high resolution model runs as well as the design for the low resolution model runs added, and how they are related to each other will have a strong impact on the value of the additional low resolution model runs.
    \item The particulars of the model being studied, especially how resolution interacts with parameter interpretation. For instance, our study suggests that river width error may have a more consistent interpretation across different resolutions of LISFLOOD, while the interpretation of channel roughness appears to be less consistent across resolutions.
\end{enumerate}
These are difficult variables to study -- and surely many conclusions will be situation specific -- we do not claim this study has any guidance for how to optimally design model runs, nor when a multiresolution calibration approach is preferable to a single (high resolution) calibration approach. We hope that our study will cause other researchers to consider the many challenges and tradeoffs involved in calibrating models based on low and high resolution model runs.

\section*{Supporting Information}
Additional information and supporting material for this article is available online at the journal's website.

\section*{Acknowledgements}
This work was supported by the U.S. Department of Energy, Office of Science, Biological and Environmental Research Program, Earth and Environmental Systems Modeling, MultiSector Dynamics under cooperative agreement DE-SC005171. Additional support was provided by the Penn State Center for Climate Risk Management and the Thayer School of Engineering at Dartmouth College. Any opinions, findings, and conclusions or recommendations expressed in this material are those of the authors and do not necessarily reflect the views of the US Department of Energy or other funding entities. We are very grateful to \citep{jck_ml} for the Stan \citep{stan} and R \citep{R} codes which they kindly made available on their GitHub page: https://github.com/jcken95/sml-athena. Computations for this research were performed on the Pennsylvania State University’s Institute for Computational and Data Sciences’ Roar supercomputer.

\section*{Author contributions}

\noindent All authors co-designed the overall study and took part in writing the manuscript. S.R. adapted codes for emulation, wrote codes for data processing, calibration, performance evaluation, data visualization, and wrote the initial draft of the manuscript. S.R., M.H., and B.S.L. developed the statistical methodology. M.H., and B.S.L. also assisted in writing-review and editing the manuscript. S.S. assisted in hydraulic modeling, writing-review and editing the manuscript. I.S. configured the LISFLOOD-FP hydraulic model for the given location and parameter settings and assisted in writing-review and editing the manuscript. K.K. secured funding and assisted in writing-review and editing the manuscript.

\section*{Code and Data Availability}

All codes and data are available on Github: https://github.com/samantha-roth/floodmodelcalibration.

\newpage

\bibliography{references}

\end{document}